\documentclass{aa}
\usepackage{txfonts}
\usepackage{graphicx}
\usepackage{mathrsfs}
\usepackage[usenames]{color}

\newcommand{\ud}{\mathrm{d}} 
\newcommand{\be}{\begin{equation}}
\newcommand{\ee}{\end{equation}}
\newcommand{\ds}{\displaystyle}

\begin{document}

\title{Effects of magnetic fields on the cosmic-ray ionization \\ of molecular cloud cores}

\author{Marco Padovani\inst{1,2} and Daniele Galli\inst{2}}

\institute{
Institut de Ci\`encies de l'Espai (CSIC--IEEC), 
Campus UAB, Facultat de Ci\`encies, Torre C5--parell 2$^{a}$, 
E--08193 Bellaterra, Spain\\
\email{padovani@ieec.uab.es}
\and 
INAF--Osservatorio Astrofisico di Arcetri, Largo E. Fermi 5, I--50125 
Firenze, Italy\\
\email{galli@arcetri.astro.it}
}


\abstract 
{Low-energy cosmic rays are the dominant source of ionization for
molecular cloud cores. The ionization fraction, in turn, controls the
coupling of the magnetic field to the gas and hence the dynamical 
evolution of the cores.}
{The purpose of this work is to compute the attenuation of the 
cosmic-ray flux rate in a cloud core taking into account 
magnetic focusing, magnetic mirroring, and all relevant energy loss processes.}
{We adopt a standard cloud model characterized by a mass-to-flux ratio
supercritical by a factor of $\sim 2$ to describe the density and
magnetic field distribution of a low-mass starless core, and we follow
the propagation of cosmic rays through the core along flux tubes
enclosing different amount of mass. We then extend our analysis 
to cores with different mass-to-flux ratios.}
{We find that mirroring always dominates over focusing, implying
a reduction of the cosmic-ray ionization rate by a factor of $\sim 2$--3
over most of a solar-mass core with respect to the value in the
intercloud medium outside the core. For flux tubes enclosing larger masses the reduction
factor is smaller, since the field becomes increasingly uniform at
larger radii and lower densities. We also find that the cosmic-ray ionization rate is
further reduced in clouds with stronger magnetic field, e.g. by a
factor $\sim 4$ for a marginally critical cloud.}
{The magnetic field threading molecular cloud cores affects the penetration
of low-energy cosmic rays and reduces the ionization rate by a factor
3--4 depending on the position inside the core and the magnetization of 
the core.}


\maketitle

\section{Introduction}
\label{intro}

Cosmic rays (CRs) are a primary source of ionization in interstellar
clouds (Hayakawa et al.~1961, Spitzer \& Tomasko~1968), contribute to
the heating of the gas (Glassgold \& Langer~1973) and affect the
chemistry of dense clouds by removing molecules from the most volatiles
ices like CO (Hasegawa \& Herbst 1993). For a review of CR induced
chemistry in the interstellar medium (ISM), see Dalgarno~(2006).  In
addition, the observed diffuse gamma-ray emission from the galactic
plane is believed to be the result of the decay of neutral pions
produced during inelastic collisions of high-energy ($>1$~GeV) CRs with
the ISM (see e.g. Gabici et al.~2007).

CR protons and heavy nuclei in the range 1~MeV/nucleon--1~GeV/nucleon
and CR electrons in the energy range 10~keV--10~MeV (hereafter,
low-energy CRs) are of particular interest for studies of interstellar
chemistry, as they provide the bulk of the ionization in molecular
cloud cores of typical H$_2$ column density $N\approx
10^{22}$~cm$^{-2}$.  Consequently, low-energy CRs determine the degree
of coupling of the magnetic field in the gas and control the dynamical
evolution of molecular cloud cores.  Values of the ionization fraction
in molecular cloud cores determined from the column densities of
species sensitive to the electron density extend over about 1--2 orders
of magnitude (Caselli et al.~1998).  It is unclear whether the inferred
CR ionization rates span a comparable range or not, due to the
sensitivity of the results to the adopted chemical model (Williams et
al.~1998, Wakelam et al.~2006). Since the scatter in the ionization
fraction may, in part, reflect intrinsic variations of the CR flux from
core to core, it is important to assess the question of the penetration
of CRs in molecular cores.

In a previous study, Padovani et al.~(2009, hereafter P09) computed the decrease of
the CR ionization rate as function of column density in a
plane-parallel, unmagnetized cloud, considering energy losses due to
elastic and non-elastic collisions of CRs with the ambient gas,
bremsstrahlung, and pion production.  However, interstellar magnetic
fields can affect the propagation of CRs by magnetic focusing,
mirroring, and diffusion (Skilling \& Strong~1976, Cesarsky \&
V\"olk~1978, Chandran~2000, Padoan \& Scalo~2005). The first two
effects are due to the non-uniformity of the large-scale (mean)
component of the field, whereas the latter is associated to magnetic
field fluctuations on the scale of the Larmor radius of CR particles.
The relative importance of these processes in the ISM depends on a
number of variables not always well determined (geometry and strength
of the magnetic field, nature and characteristics of turbulence, etc.),
and is unclear whether they can significantly reduce (or enhance) the
CR ionization of molecular cloud cores. In addition, the damping of
small-scale magnetic fluctuations (e.g. Alfv\'en waves) that affect the
propagation of CRs is strongly dependent on the ionization fraction of
the medium, which, in turn, is mostly determined by the CRs
themselves. The general problem must then be addressed in a
self-consistent way.

In this paper we consider the propagation of CRs along the magnetic
field lines threading a molecular core assuming that only the mean
field affects the CR properties through a combination of magnetic
focusing and mirroring.  A similar problem was addressed by Desch et
al.~(2004) to determine the efficiency of spallation reactions and
trapping of CR nuclei in the protosolar nebula. In their study, Desch
et al.~(2004) computed an energy dependent correction factor for the CR
flux, representing the fraction of the core's column density a CR of
energy $E$ interacts with as it passes through the core or it is
thermalized.  Our approach is similar, but we extend and refine their
method, incorporating the magnetic effects in the propagation model of
P09, that  takes into account energy losses, and
adopting a specific profile for the density and magnetic field of the
core.

This paper is organized as follows: in Sect.~\ref{CRinteractionB} we
discuss the effects of the magnetic field on the CR propagation; in
Sect.~\ref{cloudmodel} we describe the adopted cloud core model; in
Sect.~\ref{ionization} we compute the modifications of the CR
ionization rate due to the magnetic field; in Sect.~\ref{dependence} we
explore the dependence of the results on the mass and magnetization of
the core; finally, in Sect.~\ref{conclusions} we summarize our
conclusions.

\section{Effects of magnetic fields on CR propagation}
\label{CRinteractionB}

CRs are charged particles that perform helicoidal trajectories around
the magnetic field lines of the medium where they propagate. For a
particle of charge $Ze$ and mass $m$ in an uniform magnetic field, this
motion consists of the combination of a circular motion around the
field line with velocity $v_\perp$ and a uniform motion along the field
line with velocity $v_\parallel$.  The frequency of the circular motion
is the {\em cyclotron frequency}, $\Omega_{\rm c}=ZeB/\gamma mc$. The
Larmor radius $r_{\rm L}$ is given by
\be
r_{\rm L}=\frac{v_\perp}{|\Omega_{\rm c}|}=
\frac{pc}{ZeB}\sin\alpha\;,
\ee
where $p=\gamma m v$ is the particle's momentum and $\alpha$ is the {\em pitch
angle} (i.e. the angle between the particle's velocity and the magnetic
field). The {\em helical step} $\delta$, that is the projection along
the direction of the magnetic field of the pathlength of the CR during
a single rotation around the field lines, is given by
\be
\delta=v_\parallel \frac{2\pi}{|\Omega_{\rm c}|}=
2\pi\frac{pc}{ZeB}\cos\alpha\,.
\ee
For a cloud with a magnetic field $B=10$~$\mu$G, the Larmor
radii of ionizing CRs (CR protons and heavy nuclei with $E \lesssim
1$~GeV/nucleon and CR electrons with $E \lesssim 10$~MeV, see P09) are
less than $\sim 10^{-7}$~pc and $\sim 10^{-9}$~pc for protons and
electrons, respectively, many orders of magnitude smaller than the
typical size of Bok globules ($\sim 0.05$~pc), dense cores
($\sim$ 1--5~pc), and giant molecular clouds ($\sim 25$~pc). 
In the absence of small-scale
perturbations in the field, we can therefore assume that low-energy CRs
propagate closely following the magnetic field lines.

Perturbations in the forms of magnetohydrodynamic (MHD) waves with
wavelength of the order of the Larmor radius of the particle can
efficiently scatter CRs.  The waves can be part of an MHD turbulent
cascade, or can be self-generated by the CRs themselves
(Kulsrud~2005).  However, in a mostly neutral ISM, turbulent MHD
cascades are quenched at scales of roughly the collision mean free path
of ions with neutrals, if the ion-neutral collision rate exceeds the
energy injection rate. Alfv\'en waves with wavelength $\lambda <
\lambda_{\rm cr}=\pi v_A/(\gamma_{in}\rho_i)$, where $v_A$ is the
Alfv\'en speed in the neutrals and $\gamma_{in}$ the collisional drag
coefficient, are efficiently damped by collisions with neutrals
(Kulsrud \& Pearce~1969). For typical molecular cloud conditions
($n=10^4$~cm$^{-3}$, $B=10$~$\mu$G, and ionization fraction $\sim
10^{-7}$), the critical wavelength is $\lambda_{\rm cr}\approx 3\times
10^{-3}$~pc, assuming $\gamma_{in}\approx
10^{14}$~cm$^3$~s$^{-1}$~g$^{-1}$ (Pinto et al.~2008). As we have seen,
the Larmor radii of ionizing CR particles are much smaller than the
cloud's size or the critical wavelength. 
Thus, 
for typical values of the cloud's parameters, only CR particles
with energy larger than a few TeV, and Larmor radii $r_{\rm L} \ge
\lambda_{\rm cr}$ find MHD waves to resonate with. These particles 
however do not contribute significantly to the ionization of the cloud.

The origin and maintenance of MHD disturbances at the scale of the
Larmor radius is unclear: small-scale MHD waves can be self-generated
by CRs streaming in the intercloud medium (ICM) or in the cloud itself. Cesarsky \&
V\"olk~(1978) have shown that the streaming instability operates in the
ICM only for CRs with energy below $\sim 40$~MeV/nucleon (for a region
of size 1~pc, density $10^3$~cm$^{-3}$, and magnetic field 10~$\mu$G)
and at even lower energies in the regions surrounding molecular cloud
cores. Thus, we expect that CRs in the energy range between
100~MeV/nucleon and 1~GeV/nucleon, that provide the bulk of the
ionization in a cloud core, stream freely through the core without
self-generating MHD waves. We ignore therefore the presence of
self-generated waves in cloud cores for the rest of this paper, but we
believe that this problem deserves further scrutiny.

\subsection{Magnetic focusing and mirroring}
\label{focusmirror}

Theoretical models predict that collapsing cloud cores must overcome
the support provided by their magnetic field in order to form stars. In
the process, the competition between gravity pulling inward and
magnetic pressure pushing outward is expected to produce a warped, {\em
hourglass} pattern of the magnetic field. Recently, this scenario has
received support from observations.  Maps of polarized dust emission
have revealed that the magnetic field in molecular clouds is rather
uniform, except near cores where the field becomes strongly pinched and
almost radial (see, e.g., Tang et al.~2009).  On the other hand
high-resolution interferometric observations of submillimeter polarized
emission in the low-mass core NGC~1333~IRAS4A by Girart et al.~(2006)
show a magnetic field geometry consistent with the predictions of
theoretical models for the formation of solar-type stars, in which
ordered large-scale magnetic fields control the evolution and collapse
of molecular cloud cores (see a comparison of observations with
theoretical collapse models in Gon\c calves et al.~2008). We therefore
adopt the hourglass geometry as the basis of our analysis of CR
penetration into a cloud core (see Sect.~\ref{cloudmodel}).

The effects of magnetic mirroring and focusing in a hourglass geometry
can be simply described following e.g. Desch et al.~(2004).  A charged
particle traveling in a magnetized medium conserves its kinetic energy
$\gamma mc^2$ and its magnetic moment $\mu=\gamma
mv^2\sin^2\alpha/2B$.  It follows that CRs propagating from the ICM to
the cloud's interior must increase $v_\perp^2$ to conserve $\mu$ and
decrease $v_\parallel$ to conserve $|{\bf v}|^2$. Thus, the pitch angle
of the particle must increase from the value $\alpha_{\rm ICM}$ to a
value $\alpha$ as
\be
\frac{\sin^2\alpha}{\sin^2\alpha_{\rm ICM}}=\frac{B}{B_{\rm ICM}}\equiv \chi\,,
\label{pitch}
\ee
where $\chi>1$. Therefore, a CR starting in the ICM with a pitch angle
$\sin\alpha_{\rm ICM} > 1/\chi^{1/2}$ cannot penetrate a region with magnetic
field $B >\chi B_{\rm ICM}$, and will be bounced out ({\em magnetic
mirroring}).  Conversely, the CR flux $j(E)$ in the cloud is increased
by the opening out of the field lines by a factor proportional to the
density of magnetic field lines per unit area ({\em magnetic
focusing}),
\be
j(E)=\chi j_{\raisebox{-3pt}{\tiny\rm{ICM}}}(E)\,.
\ee
The effects of focusing and mirroring depend only on the magnetic
field strength, and are the same for CR protons, electrons, and 
heavy nuclei. 

\section{The cloud model}
\label{cloudmodel}
\label{nonuni}

\begin{figure}[t]
\begin{center}
\includegraphics[scale=1]{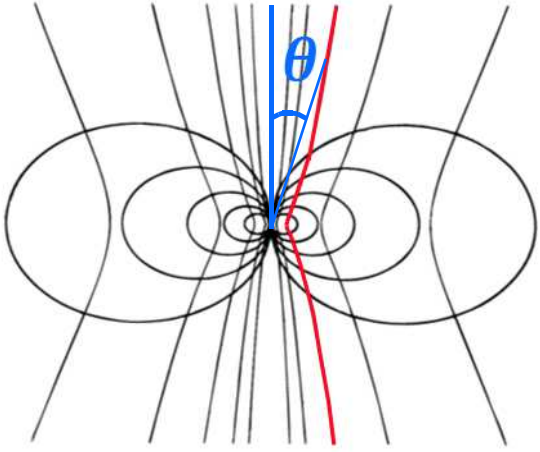}
\caption[]{Isodensity contours ({\em thick curves}) and magnetic field
lines ({\em thin curves}) in the meridional plane of a model with
$\lambda=2.66$ ($\lambda_r=1.94)$, from Li \& Shu~(1996). We consider
the motion of CRs along a particular field line ({\em red line}) as 
function of the polar angle $\theta$.}
\label{16853fg1}
\end{center}
\end{figure}

In order to study the effects of the magnetic field on the propagation
of CRs in molecular cores, we adopt the models of Li \& Shu~(1996)
(see also Galli et al.~1999), for magnetostatic, scale-free,
self-gravitating clouds supported by axially-symmetric hourglass-like
magnetic fields (see Figure~\ref{16853fg1}).  For simplicity, we consider
models with an isothermal equation of state.  These models are
characterized by a value of the non-dimensional mass-to-flux ratio
$\lambda$ defined by
\be
\lambda=2\pi G^{1/2}\frac{M(\Phi)}{\Phi}\,,
\label{def_lambda}
\ee
where $G$ is the gravitational constant, $\Phi$ the magnetic flux, and
$M(\Phi)$ the  mass contained in the flux tube $\Phi$.  A related
quantity is the {\it spherical} mass-to-flux ratio $\lambda_r<\lambda$,
evaluated from Eq.~(\ref{def_lambda}) but considering only the mass
enclosed in a sphere tangent to the ``waist'' of the flux tube.  The
latter quantity is more appropriate for comparisons with observations,
because of the limited beam of a telescope.

The magnetic field of the model is axially symmetric and purely 
poloidal. In spherical coordinates ($r, \theta$), it is given by 
\be
{\bf B}=\frac{1}{2\pi}\nabla\times\left(\frac{\Phi}{r \sin\theta}\,
\hat{\mathbf{e}}_\phi\right)\,,
\label{defB}
\ee
where 
\be
\Phi(r,\theta)=\ds{\frac{4\pi c_s^2r}{G^{1/2}}}\phi(\theta)\,,
\label{Phi}
\ee
$c_s$ is the sound speed, and $\phi(\theta)$ is a dimensionless
function. Similarly, the density is given by
\be
\rho(r,\theta)=\ds{\frac{c_s^2}{2\pi Gr^2}}R(\theta)\,, 
\label{rho}
\ee
where $R(\theta)$ is a dimensionless function.  From
Eq.~(\ref{defB}) and (\ref{Phi}), the radial and polar components of
the magnetic field are respectively
\be
B_r=\frac{2c_s^2}{G^{1/2}r\sin\theta}\phi^\prime 
\qquad\mbox{and}\qquad
B_\theta=-\frac{2c_s^2}{G^{1/2}r\sin\theta}\phi\,,
\ee
where $\phi^\prime={\rm d}\phi/{\rm d}\theta$. 
The modulus of the magnetic field vector is
\be
\label{modB}
|{\bf B}| = \frac{4\lambda c_s^{4}\phi^2}{G^{3/2}M(\Phi)\sin\theta}
\sqrt{1+\left(\frac{\;\phi^\prime}{\phi}\right)^2}+B_{\rm ICM}\,,
\ee
where $B_{\rm ICM}$ is the intercloud magnetic field, assumed uniform
and equal in strength to the Galactic magnetic field.

\begin{figure}[t]
\begin{center}
\includegraphics[scale=1]{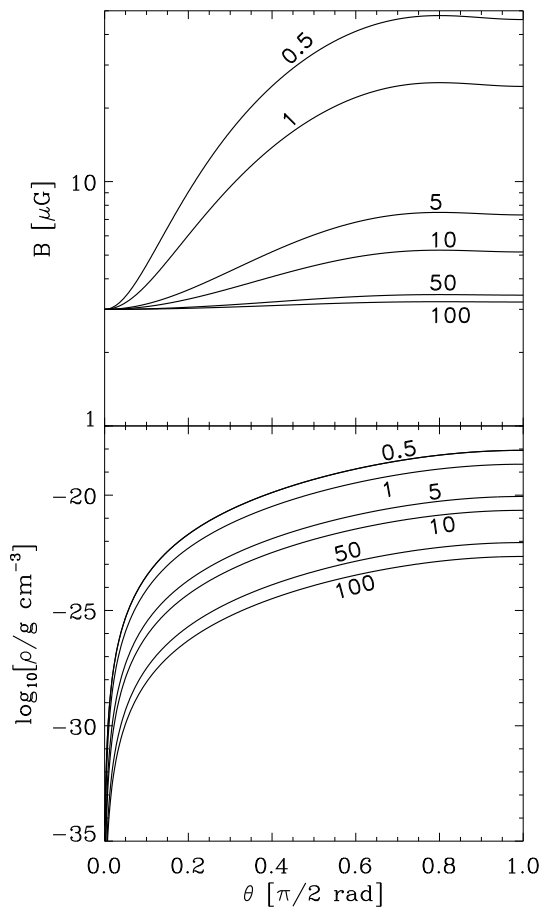}
\caption[]{Modulus of the magnetic field ({\em upper panel}) and
density ({\em lower panel}) as function of the polar angle $\theta$ for
flux tubes enclosing different masses (label values in $M_\odot$).}
\label{16853fg2}
\end{center}
\end{figure}

For our reference model we choose $\lambda=2.66$ ($\lambda_r=1.94$),
and we assume $c_s=0.2$~km~s$^{-1}$, $B_{\rm ICM}=3$~$\mu$G.
Figure~\ref{16853fg2} shows the profiles of the magnetic field
strength and the density as function of the polar angle, moving along
flux tubes enclosing various masses.  In the following we will focus on
a flux tube enclosing a mass $M(\Phi)=1$~$M_\odot$, a typical value for
a low-mass core, corresponding to an equatorial radius at the ``waist''
of the flux tube $r_{\rm eq}=0.036$~pc (other cases will be considered
in Sect.~\ref{dependence}).

\begin{figure}[t]
\begin{center}
\includegraphics[scale=1]{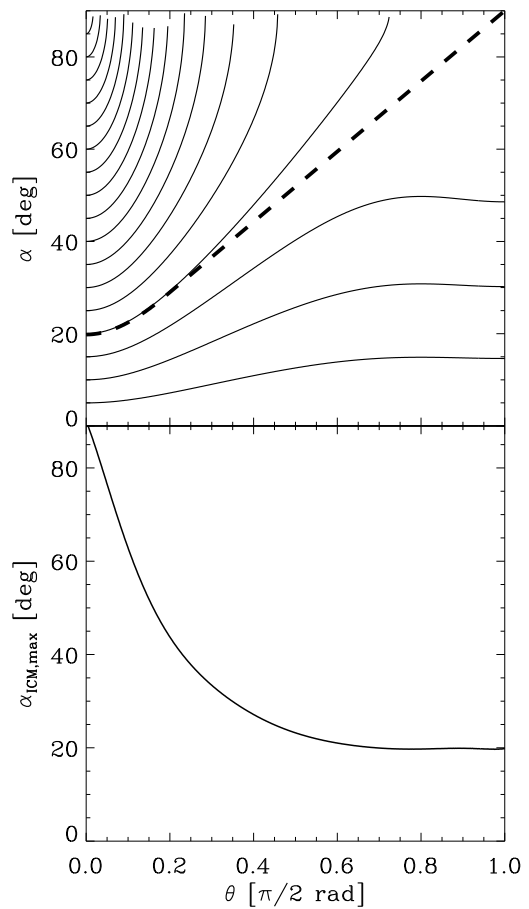}
\caption[]{{\em Upper panel}: variation of the CR pitch angle $\alpha$
as function of the polar angle $\theta$ for values of $\alpha_{\rm
ICM}$ between $0^\circ$ and $90^\circ$ in steps of $5^\circ$ (from bottom to top); 
the {\em dashed line} represents the critical pitch angle $\alpha_{\rm
cr}=20.5^\circ$. {\em Lower panel}: relation between
the maximum initial pitch angle $\alpha_{\rm ICM, max}$ and the polar
angle that can be reached during the propagation inside the core.}
\label{16853fg3}
\end{center}
\end{figure}

The upper panel of Figure~\ref{16853fg3} shows the variation of the
pitch angle $\alpha$ computed from Eq.~(\ref{pitch}) as function of the
polar angle $\theta$ for various inital values of $\alpha_{\rm ICM}$
assuming the magnetic field profile of the reference flux tube (see
Figure~\ref{16853fg2}).  In this case, only particles starting
with pitch angle $\alpha_{\rm ICM}<\alpha_{\rm cr}=20.5^\circ$ are
able to reach the cloud's midplane ($\theta=\pi/2$, where
$\chi=8.174$). CRs starting with pitch angles larger than this value
will be pushed back by magnetic mirroring before reaching the midplane
at a position $\theta_{\rm max}(\alpha_{\rm ICM})<\pi/2$.  Inverting
this relation, one finds the value of the maximum allowed pitch angle
$\alpha_{\rm ICM, max}(\theta)$ for a CR to reach a given position
$\theta$, shown in the lower panel of Figure~\ref{16853fg3}.

\section{CR ionization rate}
\label{ionization}

Consider first CRs entering the core from above (in the following
denoted by a subscript $+$). Only those with pitch angle $\alpha_{\rm
ICM}<\alpha_{\rm ICM,max}(\theta)$ can reach a position $\theta$ inside
the cloud. Thus, their contribution to the CR ionization rate of H$_2$ is 
\begin{eqnarray}
\zeta_+^{\rm H_2}(\theta) &=& 2\pi\ \chi(\theta)\int_0^\infty \ud E
\int_0^{\alpha_{\rm ICM, max}(\theta)}
j[E,N(\theta,\alpha_{\rm ICM})]\times\nonumber\\
&\times& [1+\phi(E)]\ \sigma^{\rm ion}(E)\ 
\sin\alpha_{\rm ICM}\ \ud\alpha_{\rm ICM}\,,
\label{ion1}
\end{eqnarray}
where $N(\theta,\alpha_{\rm ICM})$ is the column density of H$_2$ 
into the core (with
$N=0$ at $\theta=0$), $\phi(E)$ is a correction factor accounting
for the ionization of H$_2$ by secondary electrons, and $\sigma^{\rm ion}(E)$
is the ionization cross section of H$_2$.
We adopt the {\it continuous-slowing-down approximation} 
to relate the flux of CR particles of energy $E$ 
at a column density $N$ inside the cloud, 
$j[E,N(\theta,\alpha_{\rm ICM})]$, to the flux of CR particles of
energy $E_{\rm ICM}$ in the ICM, $j(E_{\rm ICM},0)$: 
\be
j[E,N(\theta,\alpha_{\rm ICM})]
=j(E_{\rm ICM},0)\frac{L(E_{\rm ICM})}{L(E)},
\ee
where $L(E)$ is the energy loss function (see P09 
for details on the method). The initial and 
final energies are such that 
\be\label{range}
N=n[R(E_{\rm ICM})-R(E)],
\ee
where $n$ is the local density of H$_2$ and $R(E)$ is the
particle's {\it range}.  We consider CRs composed of electrons, protons,
and heavy nuclei. As stressed by Webber~(1998) and P09, CR electrons
contribute significantly to the total ionization rate, and cannot be
neglected. We adopt four combinations of proton/electron spectra
$j(E)$, labeled as in P09:  Webber~(1998, W98) and
Moskalenko et al.~(2002, M02) for protons and heavy nuclei; Strong et
al.~(2000, C00 and E00) for electrons. The spectra M02 and E00 are
characterized by an increase at low energies, whereas the spectra W98
and C00 are constant or slightly decreasing.

\begin{figure}[t]
\begin{center}
\includegraphics[scale=0.6]{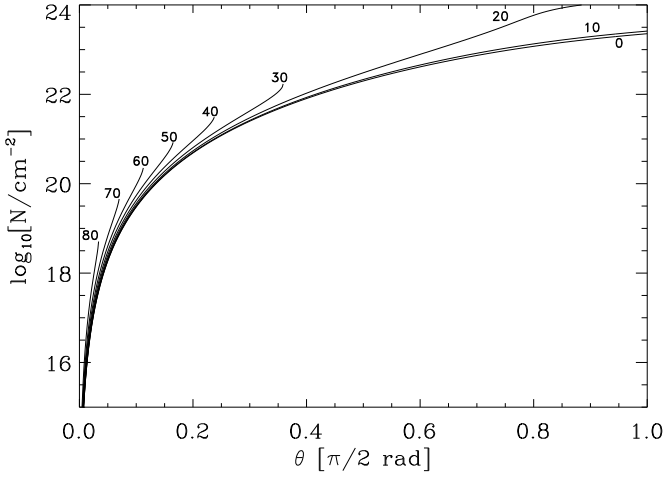}
\caption[]{Column density passed through before reaching the mirror point by
CRs propagating along a field line of the reference flux tube enclosing
1~$M_\odot$, as function of the polar angle $\theta$ for different
values of the initial pitch angle $\alpha_{\rm ICM}$ (labels in
degrees).}
\label{16853fg4}
\end{center}
\end{figure}

The column density passed through by a CR propagating along a magnetic
field line is
\be
\label{colden}
N(\theta,\alpha_{\rm ICM})=\frac{1}{\mu\,m_{\rm H}}\int\rho\,
\frac{\ud\ell}{\cos\alpha},
\ee
where $\mu=2.8$ is the molecular weight, $\ud\ell=[\ud
r^2+(r\ud\theta)^2]^{1/2}$ is the element of magnetic 
field line, and the factor
$1/\cos\alpha$ accounts for the increase of the actual path length of a
CR performing a helicoidal trajectory around a magnetic field with
respect to the displacement along the field line. 
Substituting in Eq.~(\ref{colden}) the expressions for $\rho$ and $\ud\ell$,
of our model, we obtain the actual column density passed through by CRs propagating
from the ICM to the mirror point $\theta_{\rm max}(\alpha_{\rm ICM})$,
\be
N(\theta,\alpha_{\rm ICM})=
\frac{c_s^4\lambda}{\pi \mu m_{\rm H}G^2 M(\Phi)} 
\int_0^{\theta_{\rm max}}
\frac{\phi(\theta)R(\theta)}{\cos\alpha(\theta)}
\sqrt{1+\left(\frac{\;\phi^{\prime}}{\phi}\right)
^2}\ \ud\theta\,,
\ee
where $\cos\alpha(\theta)$ is given in terms of
$\alpha_{\rm ICM}$ and $\chi$ by Eq.~(\ref{pitch}). 

The results are shown in
Figure~\ref{16853fg4} for the reference flux tube enclosing 1~$M_\odot$
and for different values of $\alpha_{\rm ICM}$.  It is evident from
the Figure that as a first approximation one can assume that
CRs coming from the ICM and traveling toward the cloud's midplane
experience a similar increase in column density, independently on the
initial pitch angle $\alpha_{\rm ICM}$, the latter mainly determining
the value of the column density at which the CRs are pushed out by
magnetic mirroring. We will then assume for all CRs the column density
profile corresponding to $\alpha_{\rm ICM}=0$, but truncated at
increasingly larger values depending on the initial pitch angle
$\alpha_{\rm ICM}$,
\be
N(\theta,\alpha_{\rm ICM}) \approx N(\theta,0) 
~~~\mbox{if~~~ $0<\theta<\theta_{\rm max}(\alpha_{\rm ICM})$}\,.
\ee
With this approximation Eq.~(\ref{ion1}) simplifies to
\begin{eqnarray}
\zeta_+^{\rm H_2}(\theta) &=& 2\pi\ \chi(\theta)
\int_0^{\alpha_{\rm ICM, max}(\theta)}\sin\alpha_{\rm ICM}
\ \ud\alpha_{\rm ICM}\times\nonumber\\
&\times&\int_0^\infty j[E,N_0(\theta)]
\ [1+\phi(E)]\ \sigma^{\rm ion}(E)\ \ud E\,,
\label{ztheta}
\end{eqnarray}
where, for simplicity, we have defined $N_0(\theta)= N(\theta,0)$ 
to indicate the column density measured along a field line.  The
second integral of Eq.~(\ref{ztheta}) is the CR ionization rate
computed assuming one-dimensional propagation, calculated in P09:
\be
\int_{0}^{\infty}j[E,N_0(\theta)]
\ [1+\phi(E)]\ \sigma^{\rm ion.}(E)\ \ud E 
\equiv \frac{1}{4\pi}\zeta_{0}^{\rm H_{2}}[N_0(\theta)]\,.
\ee
Thus, Eq.~(\ref{ion1}) can be rewritten as
\be
\label{zetal}
\zeta_+^{\rm H_2}(\theta)=\varphi_+(\theta)\ 
\zeta_0^{\rm H_2}[N_0(\theta)]\,,
\ee
where 
\be
\varphi_+(\theta)=\frac{\chi(\theta)}{2}[1-\cos\alpha_{\rm ICM,max}(\theta)]
\ee
is the factor accounting for magnetic effects.  This completes the
calculation of the contribution to the ionization of CRs coming from
the upper side of the cloud.

\begin{figure}[t]
\begin{center}
\includegraphics[scale=1]{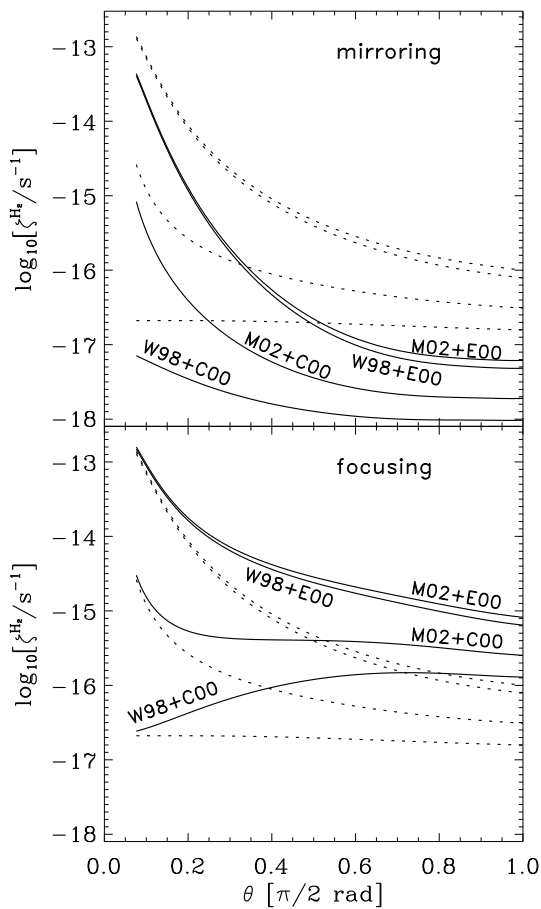}
\caption{Effects of magnetic mirroring ({\em upper panel}, {\em solid
lines}) and focusing ({\em lower panel}, {\em solid lines}) on the
CR ionization rate as function of the polar angle. {\em
Dotted lines} represent the case without considering the magnetic field
influence, from P09. Labels refer to the CR proton
and electron spectra used in P09 in the non-magnetic
field case (proton spectra:  Webber 1998, W98; Moskalenko et al. 2002,
M02; electron spectra: Strong et al. 2000, C00 and E00).}
\label{16853fg5}
\end{center}
\end{figure}

\begin{figure}[t]
\begin{center}
\includegraphics[scale=1]{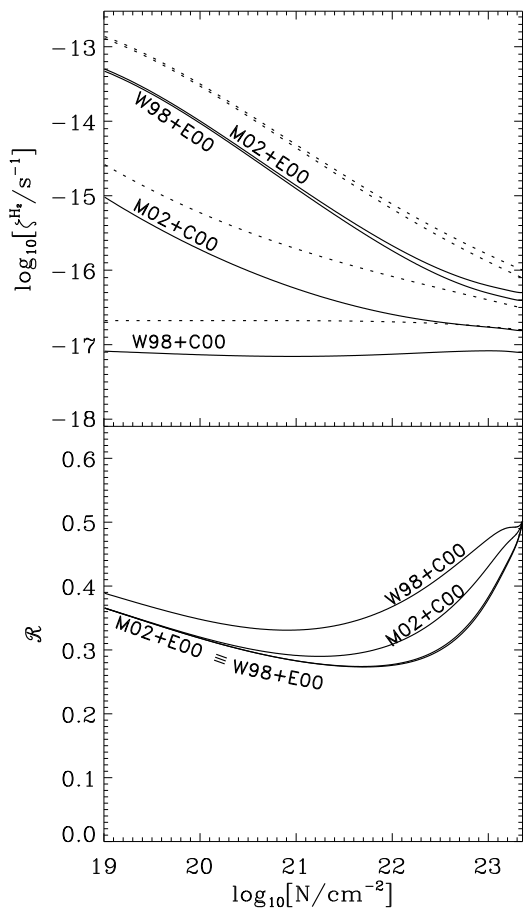}
\caption{Comparison between the CR ionization rate with and without the
effects of magnetic field ({\em solid} and {\em dotted lines},
respectively) as function of the column density ({\em upper panel});
ratio between the ionization rates in the magnetic and non-magnetic case
({\em lower panel}). The curves are labeled as in Figure~\ref{16853fg5}.}
\label{16853fg6}
\end{center}
\end{figure}

A CR propagating along a field line and reaching 
a polar angle $\theta$ coming from the lower side of the cloud 
has passed through a column density 
\be
N_0(\pi/2)+[N_0(\pi/2)-N_0(\theta)]=2N_0(\pi/2)-N_0(\theta)\,,
\ee
where $N_0(\pi/2)$ is the cloud's column density at the midplane. 
Thus, the contribution of CRs coming from the 
lower side of the cloud (denoted by a subscript $-$) to the ionization rate is 
\be
\label{zetar}
\zeta_-^{\rm H_2}(\theta)= \varphi_-(\theta)\ 
\zeta_0^{\rm H_2}[2N_0(\pi/2)-N_0(\theta)]\,,
\ee
where $\varphi_-(\theta)$ is 
\be
\varphi_-(\theta)=\frac{\chi(\theta)}{2}(1-\cos\alpha_{\rm cr})\,.
\ee
The total CR ionization rate is then given by
\be\label{zetatot}
\zeta^{\rm H_2}(\theta)=\zeta_+^{\rm H_2}(\theta)+\zeta_-^{\rm H_2}(\theta)\,.
\ee
Figure~\ref{16853fg5} shows $\zeta^{\rm H_2}(\theta)$ including the
effects of magnetic focusing and mirroring, compared to the CR
ionization rate evaluated without considering the magnetic field. The
curves refer to the four combinations of the two protons and electron
spectra assumed in this work (see the paragraph 
following Eq.~\ref{range}).  It is
evident that the two processes have opposite effects of comparable
magnitude on the ionization rate, both becoming more and more important
approaching the core's midplane where the field is stronger.  In
Figure~\ref{16853fg6} we compare the values of $\zeta^{\rm H_2}$ computed
with and without magnetic effects as function of the column density
(measured along field lines) in the core. The quantity ${\cal R}$,
defined as the ratio of the rates obtained in the two cases, is shown
in the lower panel of the Figure.  Clearly, magnetic mirroring always
reduces the CR ionization rate more than magnetic focusing can increase
it, the total effect being a net reduction of $\zeta^{\rm H_2}$ by a
factor between 2 and 3. Notice that the maximum effect of the
magnetic on $\zeta^{\rm H_2}$ is not obtained on the core's midplane,
where the field is stronger, but at an intermediate position
corresponding to column densities $10^{21}$--$10^{22}$~cm$^{-2}$.  In
fact, in the core's midplane $\varphi_+(\pi/2)=\varphi_-(\pi/2)
=\chi(\pi/2)(1-\cos\alpha_{\rm cr})/2$, and,
using Eq.~(\ref{pitch}), one has
\be
{\cal R}(\pi/2)=\chi(\pi/2)(1-\cos\alpha_{\rm cr})
=\chi(\pi/2)\left(1-\sqrt{1-\frac{1}{\chi(\pi/2)}}\right)\,
\ee
independent on the assumed CR spectrum, a result also obtained by Desch
et al.~(2004). For the reference flux tube enclosing 1~$M_\odot$,
${\cal R}(\pi/2)=0.516$. 

\section{Dependence of $\zeta^{\rm H_2}$ on 
the mass and magnetization of the core}
\label{dependence}

\subsection{Varying the enclosed mass $M(\Phi)$}

In the previous section we have considered CR propagating along the
field lines of a flux tube enclosing 1~$M_\odot$.  In
Table~\ref{tab:BvsM} we show the variation of the relevant quantities
of the problem as function of the enclosed mass for our reference core
with $\lambda=2.66$.  Flux tubes enclosing smaller masses intersect the
midplane at smaller $r_{\rm eq}$ and are characterized by larger values
of the magnetic field and density (see Figure~\ref{16853fg2}). In
particular, in our model the field strength and the density increase
inward as $|{\bf B}|\propto r^{-1} \propto M(\Phi)^{-1}$ and
$\rho\propto r^{-2}\propto M(\Phi)^{-2}$, respectively (see
Eq.~\ref{modB} and Eq.~\ref{rho}). As a consequence, focusing becomes
more important, since $\chi$ increases, but also mirroring becomes more
severe, since $\alpha_{\rm cr}$ decreases. The net effect is a stronger
reduction of $\zeta^{\rm H_2}$ in the innermost regions of the core as
compared to the envelope (see Table~\ref{tab:BvsM}). As the field
strength increases approaching the central singularity of the model,
$\chi \rightarrow \infty$ and the reduction factor 
of the CR ionization rate in the core's midplane
approaches the asymptotic value ${\cal R}(\pi/2)=1/2$.

Conversely, for flux tubes enclosing larger masses, the field strength
approaches the ICM value, and the density decreases to zero.
Therefore both focusing and mirroring become weaker, and $\alpha_{\rm
cr}$ approaches $90^\circ$, as shown by Table~\ref{tab:BvsM}.
As expected, for increasing values of $M(\Phi)$, $\zeta^{{\rm H_2}}$
approches the value of the non-magnetic case, because the magnetic
field strength decreases away from the center of the core, 
approaching the ICM value.

\begin{table}[t]
\caption{Values of the parameters described in the text as function of the 
mass, $M(\Phi)$, contained within flux tube $\Phi$.}
\begin{center}
\begin{tabular}{cccccc}
\hline\hline
$M(\Phi)$ & $r_{\rm eq}$ & $N_0(\pi/2)$ & $\alpha_{\rm cr}$ 
& $\chi(\pi/2)$ & ${\cal R}(\pi/2)$\\
$(M_\odot)$ & (pc) & (10$^{23}$ cm$^{-2}$) &  &  & \\
\hline
0.5 & 0.018 & 4.54 & 14.8$^\circ$ & 15.347 &  0.508\\
1   & 0.036 & 2.27 & 20.5$^\circ$ & 8.174  &  0.516\\
5   & 0.180 & 0.45 & 39.9$^\circ$ & 2.435  &  0.566\\
10  & 0.360 & 0.23 & 49.7$^\circ$ & 1.717  &  0.608\\
50  & 1.802 & 0.05 & 69.3$^\circ$ & 1.143  &  0.739\\
100 & 3.604 & 0.02 & 75.0$^\circ$ & 1.072  &  0.794\\
\hline
\end{tabular}
\end{center}
\label{tab:BvsM}
\end{table}%

\subsection{Varying the mass-to-flux ratio $\lambda$}

\begin{figure}[t]
\begin{center}
\includegraphics[scale=0.55]{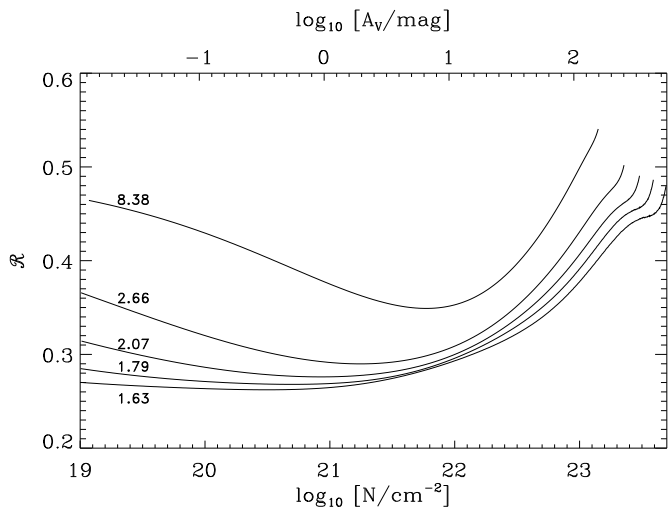}
\caption{Ratio ${\cal R}$ between the CR ionization rates in the magnetic and
non-magnetic case for the case of M02+C00 spectrum (see
Figure~\ref{16853fg5} for the labels). The curves are computed for 
a flux tube containing 1~$M_\odot$ and for different values of
$\lambda=8.38$, 2.66, 2.07, 1.79, 1.63. The upper scale shows the extinction
through the cloud obtained from $A_V/{\rm mag}=N/10^{21}~{\rm cm}^{-2}$.}
\label{16853fg7}
\end{center}
\end{figure}

In this section we explore the effects of the variation of the core's
mass-to-flux ratio $\lambda$ on $\zeta^{\rm H_2}$, for a flux tube
containing a fixed mass of 1~$M_\odot$. 
For cores with strong magnetic support (lower values of $\lambda$), the
equatorial squeezing of the field lines is stronger, and the lines
reach more internal regions of the core where the density is higher.
As a consequence, $\alpha_{\rm cr}$ decreases, the mirroring effect
becomes stronger, and a smaller fraction of CRs can penetrate the
cloud.  In Figure~\ref{16853fg7} we show the ratio ${\cal R}$ as a
function of column density for decreasing values of $\lambda$.  For
simplicity, we have considered only the combination of the proton
spectrum M02 and the electron spectrum C00 (the results obtained with
the other spectra are similar).  As the Figure shows, the
reduction of $\zeta^{\rm H_2}$ is larger in cores with larger magnetic
support, due to the increase in the field strength and concentration of
field lines. The reduction is a factor $\sim 4$ for the outer regions
of cores with $\lambda=1.63$, the lowest value of mass-to-flux 
ratio considered in our models.

For $\lambda\rightarrow 1$, the density distribution becomes more and
more flattened, the core assumes the shape of a thin disk, and the
column density from the ICM to the core's midplane becomes larger.  In
this limiting case, the magnetic field strength and the column density
increase as $(\lambda^2-1)^{-2}$. For these magnetically dominated 
disk-like configurations, the reduction of the CR ionization rate 
approaches the asymptotic value 1/2.

\section{Conclusions}
\label{conclusions}

We have determined the CR ionization rate $\zeta^{\rm H_2}$ in a
magnetically supported molecular core, and compared the results with
the case of non-magnetized density distribution with the same column
density. We have considered only the large-scale magnetic field threading
the core because, according to previous investigations, CR scattering by
magnetic fluctuation on the scale of the Larmor radius of the particle
should not be important for CRs in the energy range 100~MeV--1~GeV,
responsible for the bulk of the ionization and heating in molecular
clouds.

Assuming the hourglass magnetic field profile and the density
distribution of a model of magnetically supported cloud core, we
computed $\zeta^{\rm H_2}$ as function of position in the core, for
different values of the intercloud CR flux and composition, taking
into account the energy loss processes considered in 
the previous work of P09 and following the 
CR propagation along magnetic field lines.
The relatively weak dependence of the reduction of the CR ionization rate
on the mass enclosed in a flux tube (basically a logarithmic dependence,
see Tab.~\ref{tab:BvsM}), and on the mass-to-flux ratio of the cloud core ($\lambda
\approx 2$ on average for most cores, see Troland \& Crutcher~2008),
lends some confidence on the general validity of our results beyond the
specific cloud model adopted here. The method described in this paper
can be applied however to any axisymmetric cloud model, provided the
input quantities $B$ and $\rho$ along a given flux tube are specified
as function of the polar angle or vertical cylindrical coordinate,
as in Fig.~\ref{16853fg2}.

Our results show that $\zeta^{\rm H_2}$ in a magnetized core is always reduced
with respect to its non-magnetic value, by a factor depending on the
core's mass-to-flux ratio (determining the field strength) and the
amount of mass contained in the flux tube considered. In general, for a
flux tube enclosing about 1~$M_\odot$, and for a core with mass-to-flux
ratio $\lambda = 2.66$, the CR ionization rate is reduced by a factor of
$\sim 3$ over most of the core, and by a factor of $\sim 2$ near the core's
midplane. The reduction is less severe for flux tubes enclosing larger
masses and for larger values of the core's mass-to-flux ratio, because
of the near uniformity of the magnetic field in these cases. For cores
with mass-to-flux ratio $\lambda \approx 1.6$ or lower, the magnetic 
field can decrease the CR ionization rate by a factor of $\sim 4$ 
over most of the core's envelope.

Thus, the values of $\zeta^{\rm H_2}$ derived by Caselli et al.~(1998),
Williams et al.~(1998), Maret \& Bergin~(2007) for dense cores and
globules, probably underestimate the ``external'' (i.e. intercloud) CR
ionization rate by a factor of $\sim 3-4$, thus alleviating the
discrepancy with measurements of $\zeta^{\rm H_{2}}$ in diffuse clouds.  In this
respect, it is interesting to notice that the best overall fit with the
chemical abundances of the low-mass core TMC-1 is obtained when the CR
ionization rate is reduced by a factor $\sim 5$ with respect to the
standard value $\zeta^{\rm H_{2}}\approx 10^{-17}$~s$^{-1}$, together with
the rate of removal of atomic hydrogen from the gas phase (Flower et
al.~2007).  However, the observed large scatter in the values of
$\zeta^{\rm H_2}$ of dense clouds found by Caselli et al.~(1998), if
real, cannot be attributed to effects of the mean field or to
variations in the mass-to-flux ratio of the cores.

\acknowledgements
This work was partially supported by the Marie-Curie Research Training
Network ``Constellation'' (MRTN-CT-2006-035890).


\begin{thebibliography}{100}

\bibitem[Caselli et al.(1998)]{c98}
Caselli, P., Walmsley, C.~M., Terzieva, R. \& Herbst, E.\ 1998, \apj, 499, 234

\bibitem[Cesarsky et al.(1978)]{cv78}
Cesarsky, C.~J. \& V\"olk, H.~J.  1978, A\&A, 70, 367

\bibitem[Chandran(2000)]{c00}
Chandran, B.~D.~G.\ 2000, \apj, 529, 513

\bibitem[Dalgarno(2006)]{d06}
Dalgarno, A.\ 2006, Publ. Nat. Acad. Sci., 103, 12269

\bibitem[Desch et al.(2004)]{d04} 
Desch, S.~J., Connolly, H.~C., Jr. \& Srinivasan, G.\ 2004, \apj, 602, 528 

\bibitem[Flower et al.(2007)]{2007A&A...474..923F} Flower, D.~R.,
Pineau Des For{\^e}ts, G. \& Walmsley, C.~M.\ 2007, \aap, 474, 923 

\bibitem[Hasegawa \& Herbst(1993)]{hh93}
Hasegawa, T.~I. \& Herbst, E.\ 1993, \mnras, 263, 589 

\bibitem[Hayakawa et al.(1961)]{hnt61}
Hayakawa, S., Nishimura, S. \& Takayanagi, T.\ 1961, PASJ, 13, 184

\bibitem[Gabici et al.(2007)]{gab07} 
Gabici, S., Aharonian, F.~A. \& Blasi, P.\ 2007, \apss, 309, 365 

\bibitem[Galli et al.(1999)]{gl99}
Galli, D., Lizano, S., Li, Z.-Y., Adams, F.~C. \& Shu, F.~H.
1999, \apj, 521, 630

\bibitem[Girart et al.(2006)]{gr06}
Girart, J.~M., Rao, R. \& Marrone, D.~P.
2006, Science, 5788, 812

\bibitem[Glassgold \& Langer(1973)]{gl73}
Glassgold, A.~E. \& Langer, W.~D.\ 1973, \apj, 186, 859 

\bibitem[Gon\c calves et al.(2008)]{gg08}
Gon\c calves, J., Galli, D. \& Girart, J.~M. 2008, A\&A, 490, L39

\bibitem[Kulsrud(2005)]{k05}
Kulsrud, R.~M.\ 2005, Plasma physics for astrophysics, Princeton University Press

\bibitem[Kulsrud \& Pearce(1969)]{kp69} 
Kulsrud, R. \& Pearce, W.~P.\ 1969, \apj, 156, 445 

\bibitem[Li et al.(1996)]{ls96}
Li, Z.-Y. \& Shu, F.~H.  1996, \apj, 472, 211

\bibitem[Maret \& Bergin(2007)]{mb07}
Maret, S. \& Bergin, E.~A.\ 2007, \apj, 664, 956

\bibitem[Moskalenko et al.(2002)]{m02} 
Moskalenko, I.~V., Strong, A.~W., Ormes, J.~F. \& Potgieter, M.~S.  2002, \apj, 565, 280

\bibitem[Padoan \& Scalo(2005)]{ps05}
Padoan, P. \& Scalo, J.\ 2005, \apj, 624, L97

\bibitem[Padovani et al.(2009)]{pg09}
Padovani, M., Galli, D. \& Glassgold, A.~E. 2009, A\&A, 501, 619 (P09)

\bibitem[Pinto et al.(2008)]{pgb08} 
Pinto, C., Galli, D. \& Bacciotti, F.\ 2008, \aap, 484, 1 

\bibitem[Skilling \& Strong(1976)]{ss76}
Skilling, J. \& Strong, A.~W.\ 1976, \aap, 53, 253

\bibitem[Spitzer \& Tomasko(1968)]{st68}
Spitzer, L. \& Tomasko, M.~G.\ 1968, \apj, 152, 971

\bibitem[Strong et al.(2000)]{smr00} 
Strong, A.~W., Moskalenko, I.~V. \& Reimer, O.  2000, \apj, 537, 763 

\bibitem[Tang et al.(2009)]{t09} 
Tang, Y.-W., Ho, P.~T.~P., Koch, P.~M., 
Girart, J.~M., Lai, S.-P. \& Rao, R.\ 2009, \apj, 700, 251 

\bibitem[Troland et al.(2008)]{t08}
Troland, T. H. \& Crutcher, R. M.\ 2008, \apj, 680, 457

\bibitem[Wakelam et al.(2006)]{w06} 
Wakelam, V., Herbst, E., Selsis, F. \& Massacrier, G.\ 2006, \aap, 459, 813 

\bibitem[Williams et al.(1998)]{wal98}
Williams, J.~P., Bergin, E.~A., Caselli, P., Myers, P.~C. \& Plume, R.\ 1998, \apj, 503, 689

\bibitem[Webber(1998)]{w88}
Webber, W.~R.\ 1998, ApJ, 506, 329

\end{thebibliography}
\end{document}